\documentclass[manuscript]{aastex}

\slugcomment{Not to appear in Nonlearned J., 45.}

\shorttitle{The globular cluster AM 4}
\shortauthors{Carraro}

\begin{document}


\title{The globular cluster AM 4: yet another young globular associated
          with the Sgr Dwarf Spheroidal galaxy?}


\author{Giovanni Carraro\altaffilmark{1}}
\affil{ESO, Alonso de Cordova 3107, Vitacura, Santiago de Chile, Chile}


\altaffiltext{1}{On leave from Dipartimento di Astronomia, Universit\'a di Padova, Vicolo
             Osservatorio 3, I-35122, Padova, Italy}


\begin{abstract}
The complete census of globular clusters formerly belonging to the Sgr dSph and now
   deposited into the Galactic halo is an important contribution to our comprehension
   of the evolution and disruption of this dwarf galaxy.
    We investigate in this study the possibility that the poorly known ``old'' globular AM~4
   might be associated with the Sagittarius dwarf galaxy, and at the same time provide
   more solid estimate of its basic parameters.
   New high quality BVI photometry is presented, from which an improved Color Magnitude Diagram is
   constructed, and estimates of age and distance are then derived. The distance
  and Galactic position are finally investigated in details. 
   AM~4 is found to be a low luminosity (M$_V$=-1.82) cluster undergoing strong tidal 
   stress by the Milky Way
   and on the verge to be dissolved.
   Besides, and at odds with previous suggestions, we provide evidences that AM~4 is indeed young,
   with an age around 9 Gyrs (as Terzan~7), but somewhat more metal poor ([Fe/H=-0.97]). 
   AM~4 is located at 33$_{-4}^{+3}$ kpc from the Sun, in a direction 
   and at distance not totally incompatible with the Sgr dSph stream.
   Although we significantly improved our knowledge of AM~4, 
   further studies are encouraged to obtain radial velocity and metallicity to demonstrate
   more firmly  (or deny) the association to Sgr
\end{abstract}





\section{Introduction}

The number of star clusters associated to the Sagittarius Dwarf Spheroidal
(Sgr dSph) galaxy increased significantly in the last years due to
to the very detailed photometric and spectroscopic
studies (Mottini et al.2008, Sbordone et al. 2007, Carraro et al. 2007, Cohen 2004),
which allow to derive precise ages, distances, velocities and metallicities
of these clusters, to be compared with Sgr main body and stellar
populations (Monaco et al. 2005).
The last cluster to enter the family was Whiting~1 (Carraro et al. 2007),
but there are indications that the newly discovered Segue~1 (Belokurov et al. 2007)
might be as well associated with Sgr.\\
By increasing the number of star clusters formerly belonging to Sgr, we aim at improving our
understanding of the chemical evolution and star formation history of this dwarf.\\
\noindent
In this paper we make a new case for a possible cluster associated with 
the Sgr dSph: AM 4.\\
This loose and faint cluster was discovered by  Madore \& Arp (1982), who
placed it at the impressive distance of 200 kpc, based on the assumption that
the brightest stars in the cluster ($B \sim $21) are horizontal branch stars.\\
Later on, Inman \& Carney (1987) provided the first photometric investigation
in the B and V pass-band which showed that the brightest stars are indeed Turn
Off (TO) stars, and therefore the cluster had to be much closer to us.
Unfortunately, the data they provide are very poor, since - as the author
discuss -  the observations were hampered by unlucky conditions, and a proper
photometric calibrations was not possible. Therefore no firm conclusions
on the cluster fundamental parameters - most importantly on age and distance -
could be derived. It was only pointed out that AM~4 is probably an old and
metal poor globular.\\
Strangely enough, AM~4 remained unstudied for the last 21 years.
We made an attempt to search into various archives, especially the one hosted by 
ESO\footnote{http://archive.eso.org/eso/}, but - after proper reduction- 
did not find data of sufficient
quality to improve on Inman \& Carney (1987) early study.
This motivated the present study. 
This paper is mainly intended to draw more attention on this overlooked cluster and boost
new observational campaigns to better characterize it.\\

\noindent
The plan of this paper is as follows. In Sect.~2 we present and discuss the observational
material and the details of the reduction of the data. In Sect.~3 we describe
the new color magnitude diagram (CMD) we derived from the acquired data.
Sect.~4 is dedicate to star counts and the determination of the cluster structural properties
by using King models, whereas in Sect.~5 we derive the cluster luminosity function
and integrated magnitude. AM~4 age, reddening, metallicity and distance are discussed
in Sect.~6. Sect.~7 addresses the possibility of the membership to Srg and, finally,
Sec.~8 summarizes our results and prospects further lines of investigation.

\section{Observations and Data Reduction}
The observations were performed at 
the Las Campanas Observatory,
using the 1.0-m Swope telescope equipped with the Site$\#$3 
$2048~\times~3150$ CCD camera. The field of view is about
$14 \farcm 8~\times~22 \farcm 8$ with a scale 0.435 arcsec/pixel.
The observations were conducted the nights of  June 28,
2006. Preliminary processing of
the CCD frames was done with standard routines in the IRAF package.   
Both dome and sky flat-field frames were obtained in each
filter, and the images have also been
linearity corrected (Hamuy et al. 2006). 
The average seeing was $1\farcs20$ along the entire
night.
We observed 96 standard stars from repeated observations of the
three fields Mark A, PG 1657, PG 2213 and SA 110 (Landolt 1992)
A journal of the observations is reported in Table~1, where
a zoom of a region around AM~4 is shown in Fig.~1.\\

The following relations between the instrumental (lower case letters)  
and the standard colours and magnitudes were adopted, as derived
using 80 to 100 standard stars:

$$
V=22.115(0.004)+v-0.068(0.007) \times (B-V)+0.16(0.02) \times X
\eqno(1)
$$
$$
B=22.084(0.004)+b+0.054(0.007) \times (B-V)+0.30(0.02) \times X
\eqno(2)
$$
$$
I=22.179(0.006)+i+0.058(0.009) \times (V-I)+0.06(0.02) \times X
\eqno(3)
$$

where $X$ is an airmass. Second order color terms have been computed,
but turned out to be negligible.
The instrumental photometry was extracted
with the DAOPHOT/ALLSTAR V2.0 (Stetson 1987) package. Aperture
photometry of standards was obtained with an aperture radius
of $6\farcs69$ arcsec (14 pixels). For stars from the cluster area we
obtained profile photometry and  aperture correct
them before the transformation of instrumental photometry
into the standard system, following the techinque describe in 
Patat \& Carraro (2001, appendix A1).\\
The coordinates of all objects
were transformed to a common pixel grid defined by the reference image
using DAOMATCH/DAOMASTER.We then corrected  
the photometry for the zero-point offset in each filter and
created a master list of all objects. The instrumental magnitudes
are calculated as weighted averages of magnitudes measured
on individual frames. The final catalog contains 9477 stars and
will be made available at the CDS database.

\section{Color Magnitude Diagram}
We extracted all the stars (79 in number) within 1.8 arcmin from the cluster center and built up
the CMD shown in Fig.~2. Only the stars having photometric errors smaller than 0.1
both in color and magnitude have been considered.
The left CMD is in the V vs (B-V) plane,
whereas the right CMD is in the V vs (V-I) plane.
The two CMDs look somewhat different. The left one is quite clear, with a Turn Off Point (TO)
at V = 21.2, (B-V)= 0.4, and, in general, looks much clearer than the Inman \& Carney (1987)
one.
The sub-giant branch is as well clear but, as already emphasized by Inman \& Carney, the red Giant
Branch (RGB) is almost absent, and no hints for a Horizontal Branch (HB) or Red Giant
clump are visible. 
The Main Sequence (MS) however is well defined down to 23.5, below which our
photometry lacks completeness (see Sect.~5).
On the other hand, the right panel CMD is much sparser, and we only recognize the TO,
at  V = 21.2, (V-I) $\sim$ 0.6.
Due to the larger scatter in the star distribution in this color combination CMD,
in the following we are going to use mainly the V vs (B-V) diagram for our analysis.

\section{Cluster structure}
A glance at Fig.~1 shows clearly that AM~4 is a faint and sparse cluster lacking 
any symmetry, like Whiting~1 (Carraro et al. 2007) or Palomar 5 (Odenkirchen et al. 2001).
We performed a star count analysis to provide an estimate of the cluster structural parameter.
The center of the cluster has been chosen at RA=13:56:21.7  DEC=-27:10:03.
We constructed around this position concentric rings 15 arcsec wide to build
up a radial density profile. 

The result is shown in Fig.~3,
where the best fit King  profile is shown as solid curve, and drawn for the parameters
indicated in the upper-right corner. The symbol $\Sigma_{B}$ indicates the mean
level of the field, as counted in a region far away from the cluster, and amounting
to 40 stars/arcmin$^2$.
As already found for Palomar~13 (Cot\'e et al. 2002) and Whiting~1 (Carraro et al. 2007),
we outline the presence of a conspicuous extra-tidal star population
-in this case the tidal radius r$_t$ is at about log(r) = 2 - whose slope
in the profile is compatible with model expectations (Johnston et al 1999).
This result emphasizes how AM~4 is undergoing significant tidal stress by the
Galactic potential, which is going to lead to the complete disruption
of this cluster.
Out best fit yields a concentration parameter c=0.90, and  a core radius of about 
half an arcmin. 

\section{Luminosity Function and integrated magnitude}
With the aim of measuring the integrated apparent magnitude (V$_t$) of
AM~4 and its absolute magnitude,
we have determined its luminosity function (LF).  
To do this, we estimated the
completeness of our photometry by running experiments with artificial
stars.  We used the routine ADDSTAR
within DAOPHOT to insert 1500 artificial stars at random positions
over the whole field and
within the magnitude range of the real stars.  This number of
artificial stars was chosen so that
 a reasonable number of them (about 30\% of the real star population)
 would lie within the cluster radius
($\sim$1.8 arcmin.).  This was done for both the long and the short
 exposures.  We then reduced the images
 with the artificial stars in exactly the same manner as the real
 images.  The ratio of the number of
artificial stars recovered by ALLSTAR to the number inserted defines
the completeness.  This experiment
 was run 10 times using a different seed number with the random number
 generator that ADDSTAR uses
in the calculation of the positions and magnitudes of the artificial
stars.  The mean values that
 were found for the completeness are listed in Table~2
.
To construct the faint part of the LF, we placed two curves on either
side of the MS and parallel to the ridge line defined by the
concentration of cluster stars.
We then counted the number stars within this band that have $20.0\leq V \leq23.5
$.
In exactly the same way,
we counted the stars in the field, far beyond the cluster radius, that
lie in the same region of the CMD.
This field contribution was normalized to the area and subtracted from
each of the magnitude bins of the LF.  The background and completeness
corrected LF is shown in Fig. 4, where the error bars are the ones
indicated by Poisson statistics.
The LF keeps raising down to V $\approx$ 22 mag, then
flattens out toward faint magnitudes.  
In {\it normal} globular clusters
the LF continues to rise.  Flat LFs that resemble AM~4's have been observed
in globular clusters that
appear to be undergoing
tidal stripping by the gravitational field of the Milky Way (e.g., Pal 5, Koch et 
al. 2004; Pal 13, Cote et al. 2002; Whiting~1, Carraro et al. 2007).\\

\noindent
To obtain V$_t$, we use LF in Fig. 5. The
integration of this LF yields Vt = 15.88.
This value and the distance modulus of (m-M)$_V$ = 17.70 yield M$_V$ =
-1.82 for AM~4. We therefore confirm that AM~4 is the least luminous
globular cluster in the MW halo (Harris 1996).

\section{Cluster fundamental parameters}
Lacking any estimate of AM~4 metal abundance, we rely on the comparison with theoretical
isochrones from the Padova suite of models (Girardi et al 2000)
to derive the cluster fundamental parameters.
This is quite a difficult exercise, since the CMD does not show any obvious horizontal branch
or red clump stars, making it difficult to solve the age-metallicity degeneracy.
For this reason, we adopt a conservative approach, which consists in exploring different
ages at fixed metallicity and different metallicity at fixed age. The exercise
is shown in Fig.~5 and 6,
where we fit the star distribution in the V vs (B-V) CMD with fixed metallicity and varying
age (Fig.~5), and fixed age and varying metalliticy (Fig.~6).
The only constraints we have is the reddening, which in the direction of AM 4 is E(B-V) =0.05
according to the Schlegel et al. (1998) maps.
In. Fig~5 we explore a possible age range assuming Z = 0.001 as starting point, 
which turns into [Fe/H]=-1.27,
and adjust isochrones for ages of 8 (left panel), 9 (middle panel) and 10 Gyrs (right panel).
Looking carefully at the fit it appears that for this metallicity an age around 8-9 Gyrs provides
a better match to the TO region for an acceptable value of the reddening.\\
Turning now to Fig.~6, we try to fit the CMD assuming an age of 9 Gyrs, and  employing
different metallicity isochrones, namely Z = 0.004 (left panel), Z=0.001(middle panel) and
Z=0.0004 (right panel). This plot convincingly shows that the combination of 9 Gyr and Z=0.001
provides a good fit to the CMD, since the low metallicity isochrone implies a too large reddening
and poorly fits the TO, whereas the higher metallicity isochrone (Z=0.004) requires an unphysical
negative reddening and at the same time produces a clear mismatch in the TO region.\\
\noindent
Playing a bit more in detail with the best fit isochrone we suggest that AM~4 is 9.0$\pm$0.5
Gyrs old and have a Z=0.002 ([Fe/H = -0.97]) metal content. This final fit is shown in Fig.~7 (left panel),
whereas in the right panel we compare AM~4 with Terzan~7, which is known to be coeval (9-10 Gyrs old),
but metal richer (Z=0.004).The two clusters look almost indistinguishable in the TO region,
with some indication that AM~4 is slightly older.
The lack of an RGB in AM~4 clearly makes it difficult to compare the two clusters
in term of metallicity.
We emphasize that this results makes AM~4 much younger
that believed before (Inman \& Carney 1987) and open new horizons to understand its nature
and origin.\\
For this age and metallicity, it reddening turns out to be E(B-V)=0.04$\pm$0.01, and its apparent distance
modulus (m-M)$_V$=17.70$\pm$0.20. This implies an heliocentric distance of 33$_{-4}^{+3}$ kpc.
If we assume 8.0 kpc as the Sun distance to the Galactic center (Harris 1996), 
we find that AM~4 has the following
Galactic Cartesian coordinates: X = +21, Y = -17, and Z = +18 kpc and, hence, 
is 28 kpc far from the Galactic center. 

\section{Membership to Sagittarius?}
With these results in hands, we address now the possibility that AM~4 
is not a genuine Galaxy globular cluster,
but was formerly associated to some dwarf galaxy which deposited it into the Galactic Halo. 
The most appealing case is a possible relation with the Sgr dSph.\\
Lacking any estimate of AM~4 radial velocity, to investigate this association we make use 
of the  Law et al.  (2005) model of the Sgr stream for a spherical Galactic Dark
Matter halo, and of the Majewki et al. (2003)  
Sgr M giants sample, and compare their position and distance with AM~4 in Fig.~8.\\
\noindent
To perform this comparison, we first translate AM~4 coordinates and distance in the reference
system of the Sgr orbital plane, following Majewski et al. 
(2003\footnote{http://www.astro.virginia.edu/srm4n/Sgr}).
In the lower panel of Fig.~8, we show the position of AM~4 in the $\lambda_{\odot}$, $\delta_{\odot}$
plane. Clearly, AM~4 position differs from the M giant sample, being much higher above the Galactic plane.
Still, it is marginally compatible with the Law et al. model.\\
From the upper panel we learn on the contrary that AM~4 position is basically consistent both with
the model points and with the M giants.
This seems to indicate a possible association with Sgr, which future studies have to 
confirm or deny. \\

\section{Conclusions}
We have presented new CCD photometry in the field of the loose and faint globular
cluster AM 4, and provided improved estimates of its fundamental parameters.
The analysis of the radial density profile and LF demonstrates that AM~4 is undergoing
strong tidal interaction with the Galaxy potential, resulting in a significant loss
a low mass stars, mostly located outside the tidal radius.
This explains the loose star distribution we see in Fig.~1, which hardly resembles
a normal globular cluster. However, this star ditribution is the typical one of stripped
globular clusters, like Whiting~1, E~3 or several Palomar clusters, which reinforces the idea
of AM 4 being an almost dissolved star cluster, whose origin can indeed be extra-galactic.\\

Interestingly, we find that the cluster is younger than previously thought, with an age
close or slightly larger than Terzan~7. The metallicity we propose is not lower than [Fe/H]=-1.0,
but this value must be confirmed by future spectroscopic studies.
The young age of AM~4 
suggests that it came from the destruction of a satellite and 
the Sgr dSph is an obvious possibility.
We have shown that the cluster distance and position in the halo 
are not totally incompatible with an association
with the Sgr dSph which as well has to be more firmly investigated by measuring the radial
velocity of AM~4. 
However, both the location, and the age and metallicity we infer lend some support
to the possibilty that AM~4 was formed inside Sgr, and then deposited into the Galactic halo.

\acknowledgments

This study made use of Simbad. The author expresses his gratitude to Januzs Kaluzny
for useful insights during observations and data reduction, to Robert Zinn
for very fruitful suggestions and comments and to the referee, P. Bonifacio, for the careful
reading and important comments.

\clearpage

   \begin{figure}
   \epsscale{1.0}
   \plotone{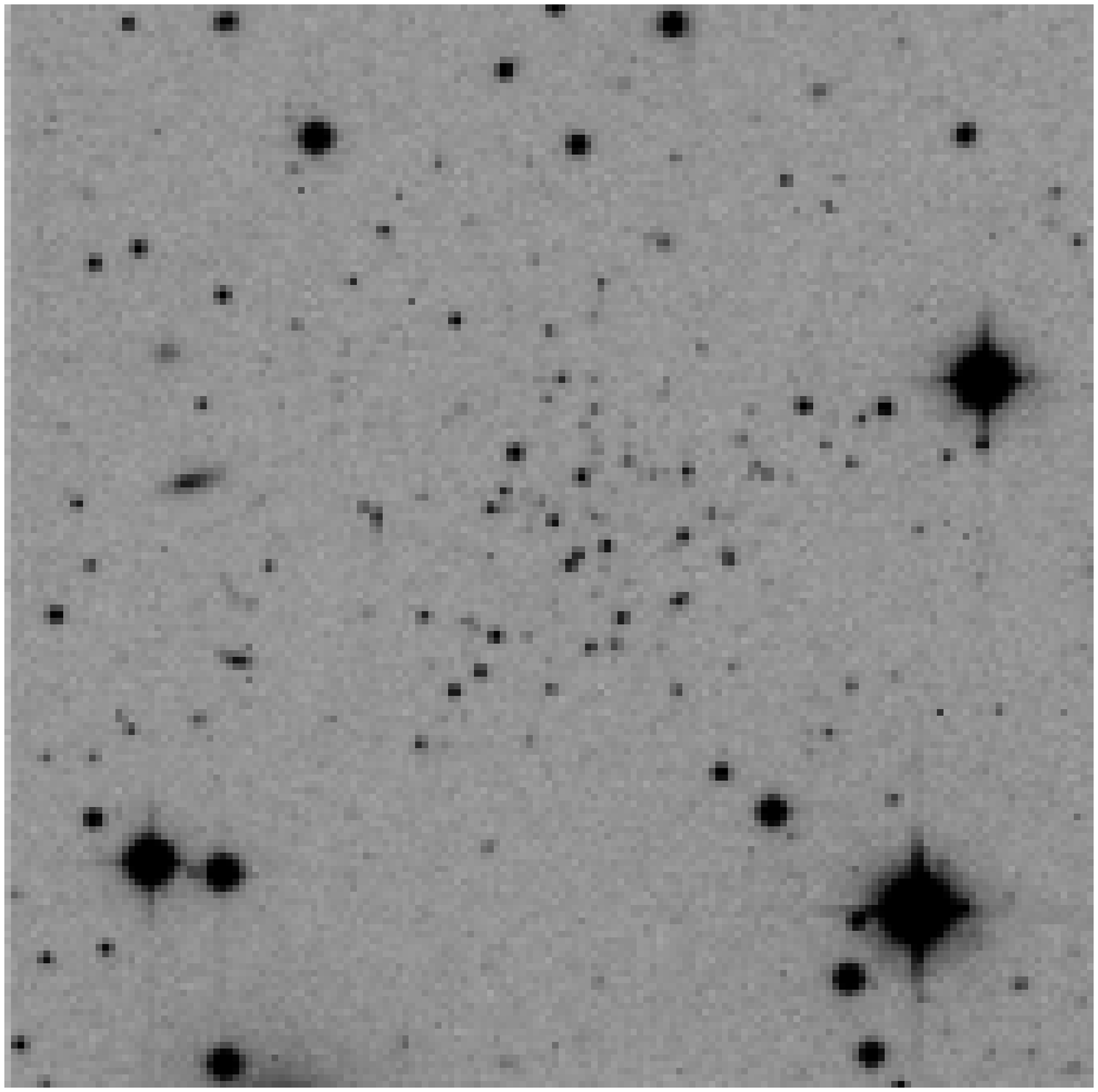}
   \caption{A zoom of a V 900 sec exposure around AM 4 region. North is up, and East
   to the left. The image is about 6 arcmin on a side.}%
    \end{figure}

\clearpage

   \begin{figure}
   \epsscale{1.0}
   \plotone{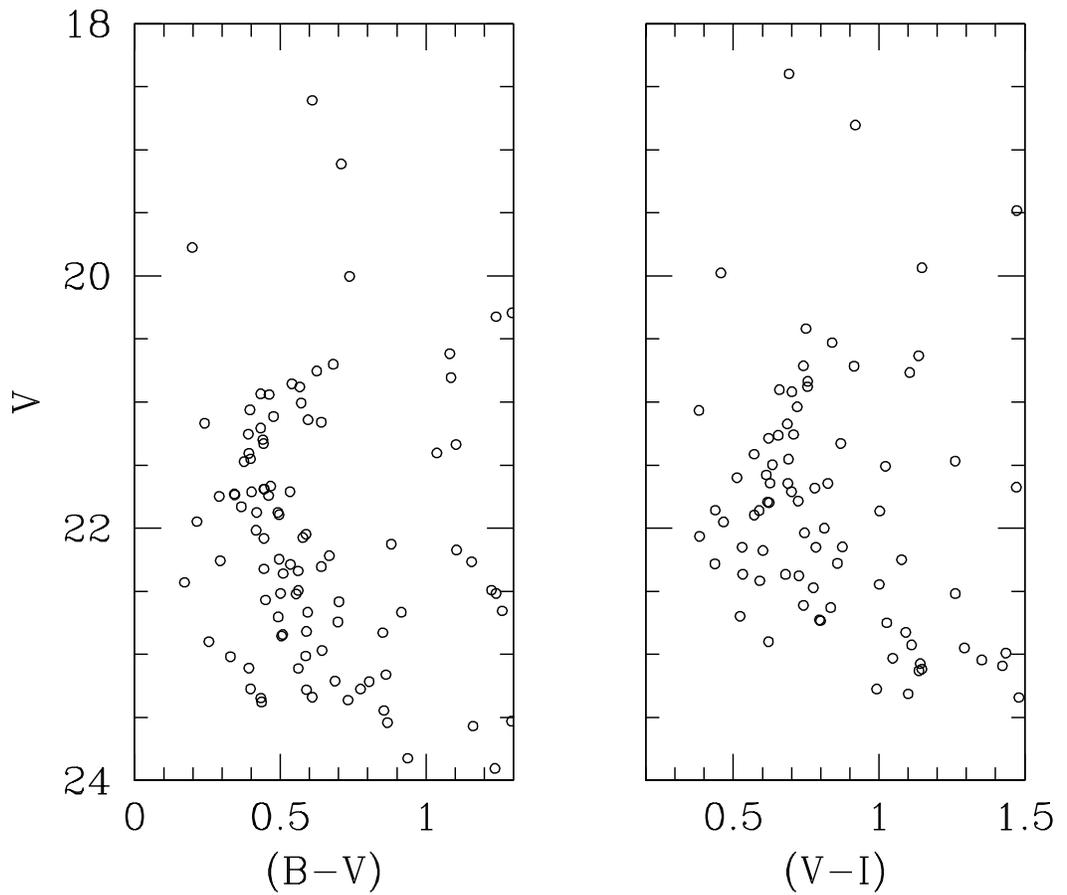}
   \caption{CMDs in the area of AM~4. Only the stars having photometric errors
    both in color and magnitude smaller than 0.1 mag, and located within 1.8 arcmin from the 
    cluster center are, considered.
    In the upper panel we show the V vs (B-V)
    CDM, while in the lower panel the V vs (V-I)}%
    \end{figure}

\clearpage

   \begin{figure}
   \epsscale{1.0}
   \plotone{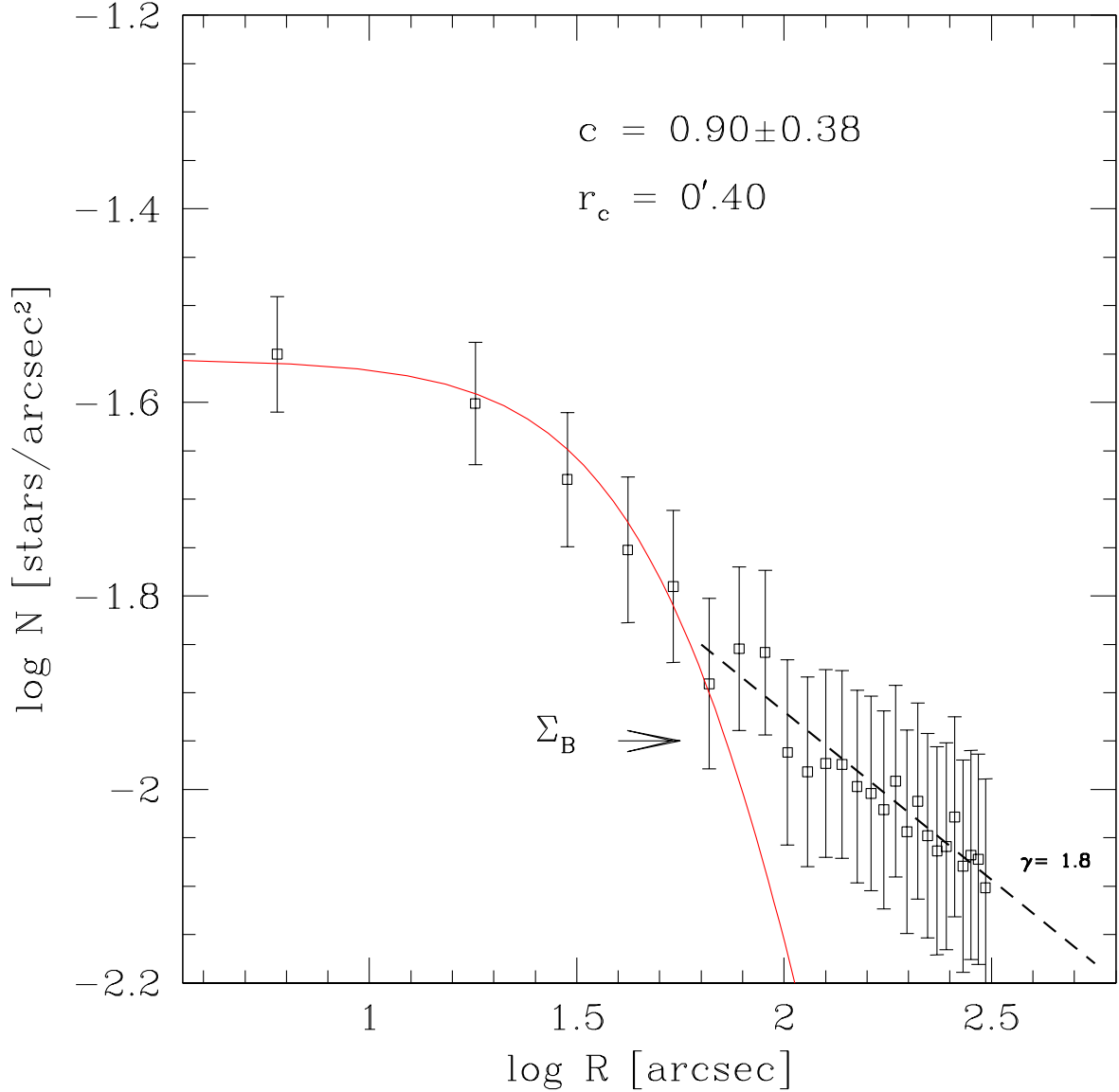}
   \caption{Surface density profile of AM~4.The solid line is the best-fit King model achieved
   with the parameters indicated in the upper-right corner. 
   Notice the strong presence of extra-tidal stars out of log(r) $\sim$ 2.
  $\Sigma_{B}$ indicates the mean  level of the field }%
    \end{figure}

\clearpage

  \begin{figure}
   \epsscale{1.0}
   \plotone{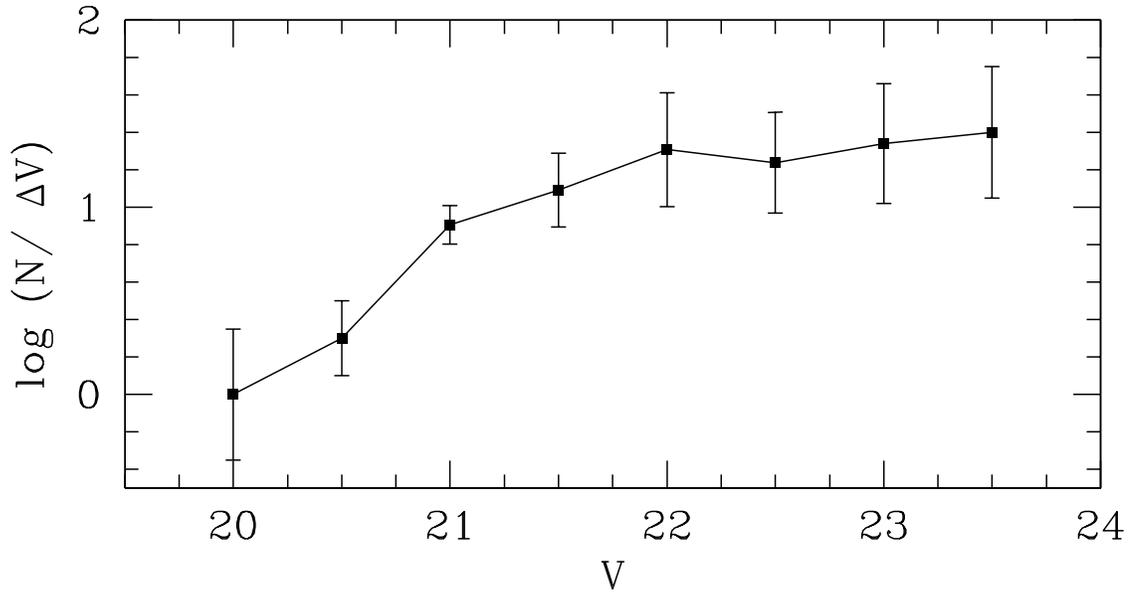}
   \caption{Completeness and field star corrected LF of AM~4}%
    \end{figure}
\clearpage

   \begin{figure*}
   \epsscale{1.0}
   \plotone{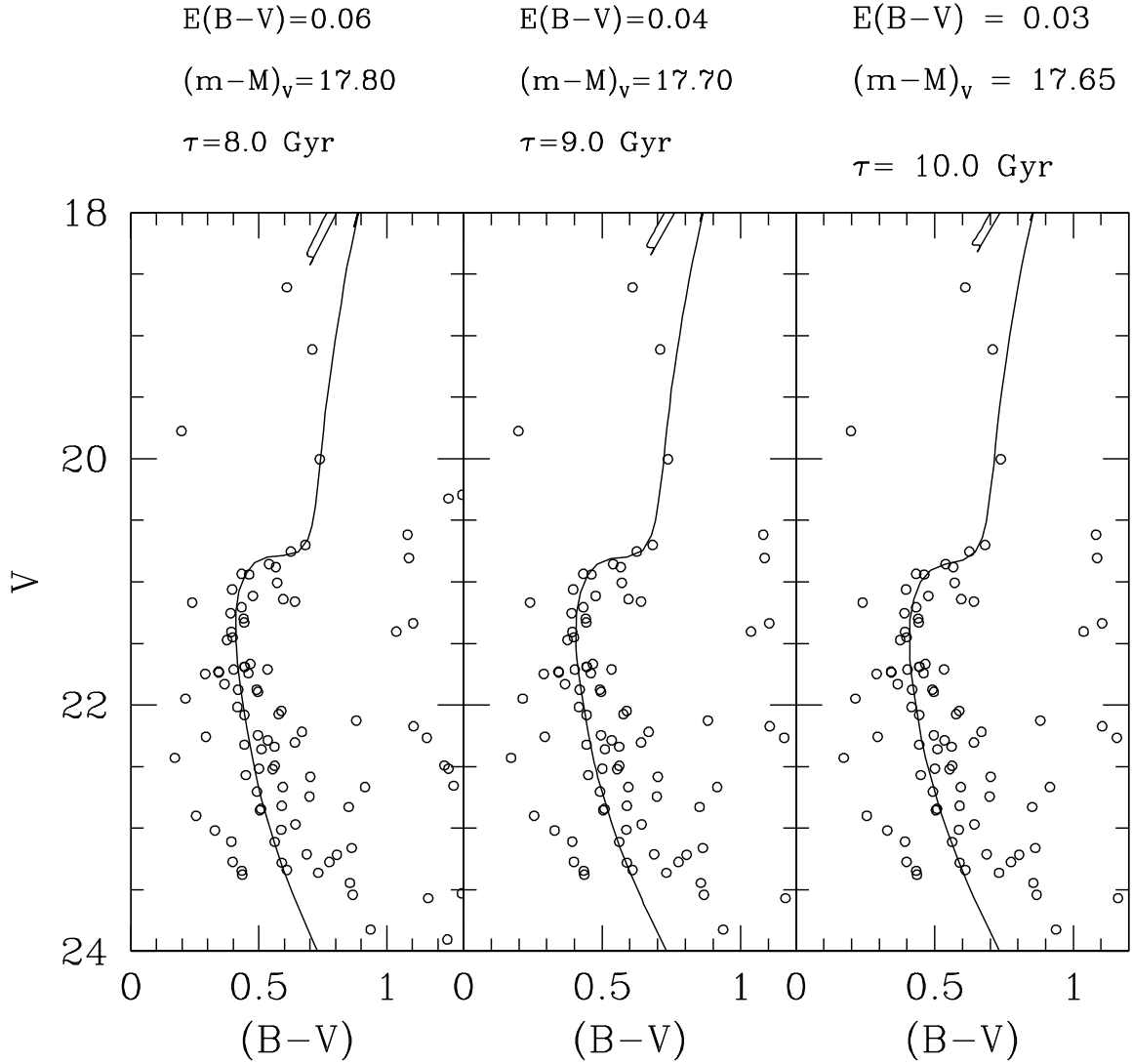}
   \caption{CMD of AM~4. Superimposed are different age isochrones adopting
    Z=0.001. The fits in each panel has been obtained for the set of parameters
    indicated above each diagram.}%
    \end{figure*}

\clearpage

   \begin{figure*}
   \epsscale{1.0}
   \plotone{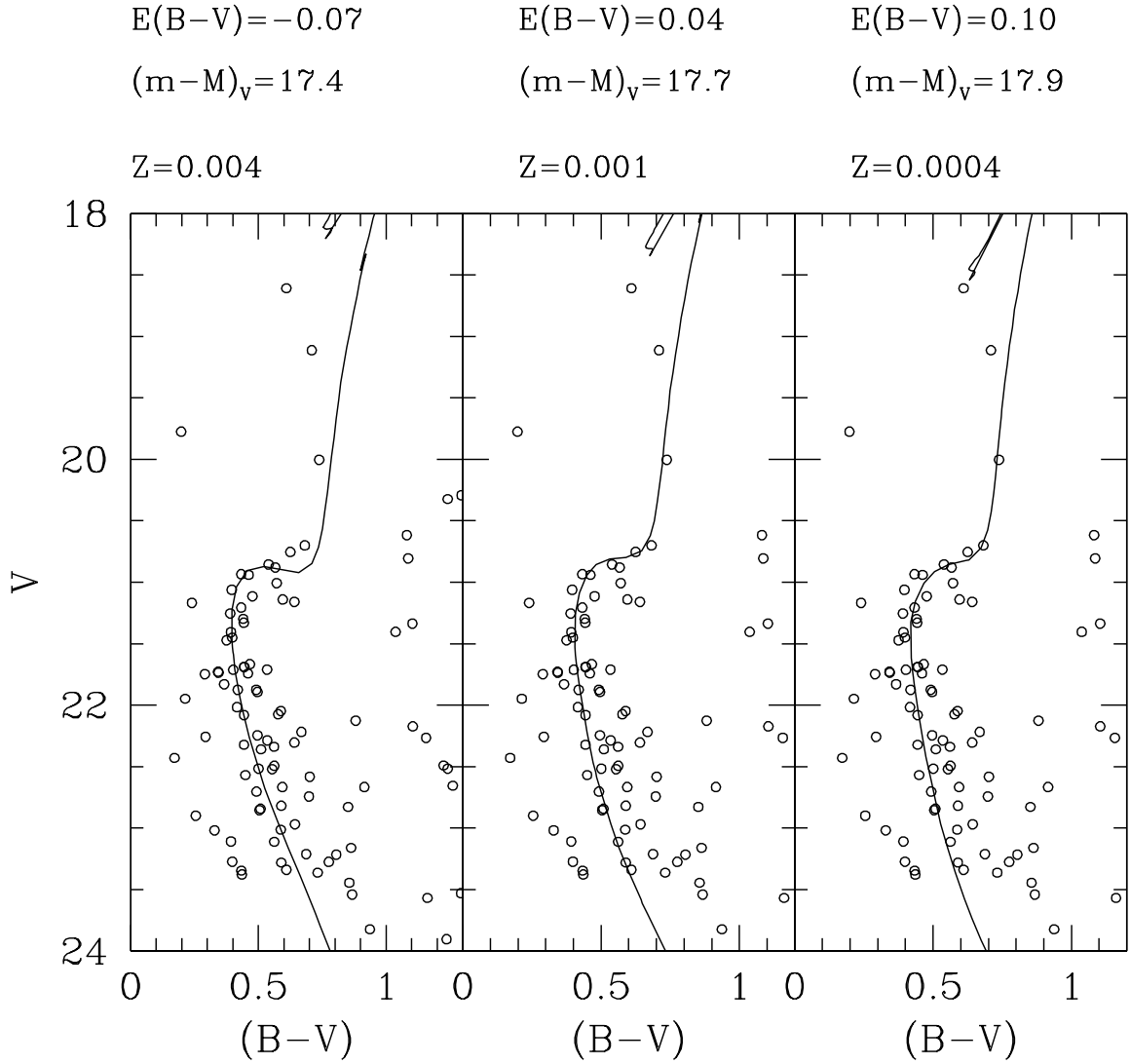}
   \caption{CMD of AM~4. Superimposed are different metallicity isochrones adopting
    an age of 9 Gyr. The fits in each panel has been obtained for the set of parameters
    indicated above each diagram}%
    \end{figure*}

\clearpage

   \begin{figure}
   \epsscale{1.0}
   \plotone{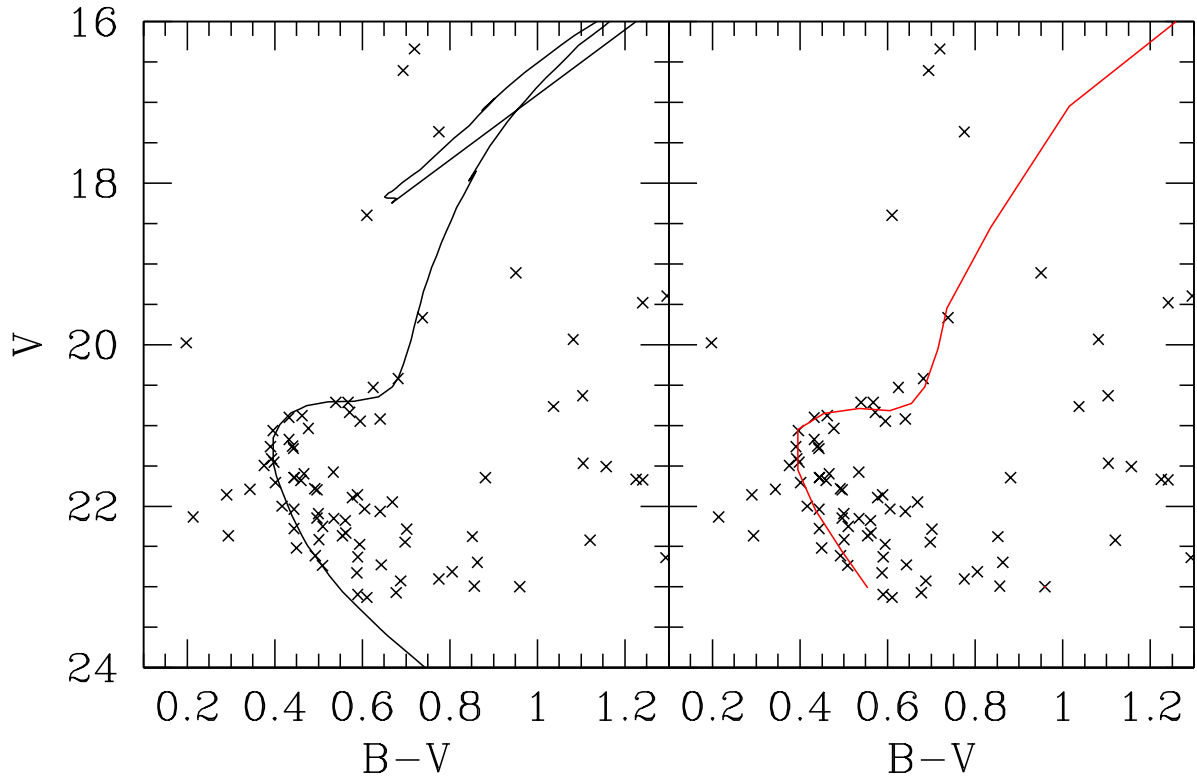}
   \caption{The best fit isochrone for an age of 9 Gyr and a metallicity Z = 0.002 is
    superimposed into AM4 CMD (left panel). The right panel presents the same AM~4 CMD,
    with superimposed Terzan~7 ridge line from Buonanno et al. 1995}%
    \end{figure}

\clearpage

   \begin{figure}
   \epsscale{1.0}
   \plotone{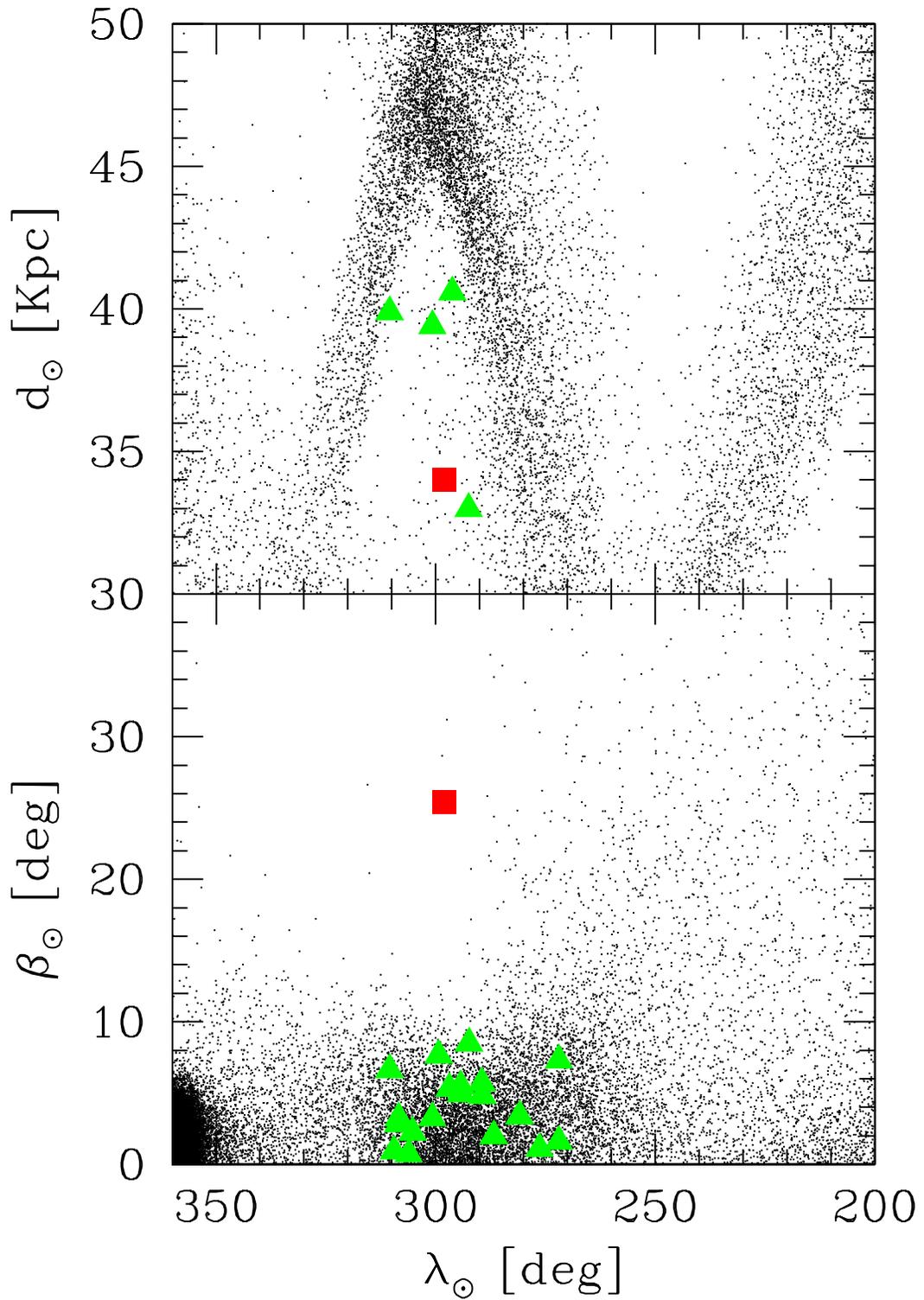}
   \caption{The Law et al. 1995 model of the Sgr tidal stream is here indicated by
    small dots. Big triangles represent Majewski et al. 2003 M giants belonging to the stream,
    and, finally, the big square stands for AM~4.}%
    \end{figure}







\clearpage

\begin{table}
\centering
\caption{Log of photometric observations on June
 28,  2006.}
\begin{tabular}{lrrr}
\hline
Cluster& Filter & Exp time (sec) & airmass\\
\hline
AM~4       & B & 15, 300, 1500, 1800  &1.00$-$1.18\\
           & V & 10, 30, 900, 1200    &1.00$-$1.07\\
           & I & 10, 30, 180, 1000     &1.01$-$1.20\\
SA110      & B & 60x2, 40x2           &1.00$-$1.45\\
           & V & 30x2, 20x3           &1.02$-$1.48\\
           & I & 30x2, 20x3           &1.10$-$1.53\\
PG~1657    & B & 30x3, 300           &1.11$-$1.40\\
           & V & 2x30, 25, 100        &1.13$-$1.46\\
           & I & 2x10, 50, 150        &1.19$-$1.55\\
Mark~A     & B & 2x10, 300            &1.00$-$1.15\\
           & V & 2x40, 50,2x100       &1.00$-$1.17\\
           & I & 2x10, 100, 2x150     &1.02$-$1.20\\           
PG~2213    & B & 2x40, 2x100,         &1.17$-$1.59\\
           & V & 2x10, 25, 2x100      &1.20$-$1.66\\
           & I & 2x10, 50, 2x150      &1.27$-$1.80\\
\hline
\end{tabular}
\end{table}
 
\clearpage

 \begin{table}
 \fontsize{8} {10pt}\selectfont
 \tabcolsep 0.1truecm
 \caption{Completeness results for AM~4}
 \begin{tabular}{cc}
 \hline
 \multicolumn{1}{c}{$V$}         &
 \multicolumn{1}{c}{$R \leq 1^{\prime}.80$}       \\
 \hline
20.5&   100.0 \\
21.0&    99.2 \\
21.5&    96.9 \\
22.0&    83.6 \\
22.5&    75.0 \\
23.0&    68.5 \\
23.5&    59.6 \\
 \hline
 \end{tabular}
\end{table}


\end{document}